\documentclass{epl}

\title{Fermionic systems with charge correlations}
\author{Ferdinando Mancini}
\institute{Dipartimento di Fisica ''E.R. Caianiello" - Unit\`{a}
di ricerca INFM di Salerno \\ Universit\`{a} degli Studi di
Salerno, I-84081 Baronissi (SA), Italy}




\begin{document}

\maketitle

\begin{abstract}
In this paper, we show that a system of localized particles,
satisfying the Fermi statistics and subject to finite-range
interactions, can be exactly solved in any dimension. In fact, in
this case it is always possible to find a finite closed set of
eigenoperators of the Hamiltonian. Then, the hierarchy of the
equations of motion for the Green's functions eventually closes and
exact expressions for them are obtained in terms of a finite number
of parameters. For example, the method is applied to the two-state
model (equivalent to the spin-1/2 Ising model) and to the
three-state model (equivalent to the extended Hubbard model in the
ionic limit or to the spin-1 Ising model). The models are exactly
solved for any dimension $d$ of the lattice. The parameters are
self-consistently determined in the case of $d=1$.
\end{abstract}

How do we study systems characterized by the presence of a huge
number of interacting particles? The standard procedure is a
perturbative one. At first, the interaction is completely ignored
or approximately taken into account as an external field (mean
field). Then, a perturbative expansion is considered. By
construction, this scheme works well only when the interaction is
sufficiently weak, but miserably fails when the correlations among
the particles are quite strong. This is the case for all new
highly interacting materials \cite{Fulde95}. An alternative
procedure has been suggested in Refs.~\cite{Mancini03,Mancini03a},
where the main ingredient is the use of composite operators. In
the presence of interactions the properties of the original (bare)
particles are modified and new particles (quasi-particles) should
be considered. The difficulty with this approach [named composite
operator method (COM)] is that the number of composite operators
rapidly increases with the number of degrees of fredom and one
must again resort to approximate treatments \cite{Mancini04}.
Recently, I have found that there is a large class of systems for
which it is always possible to find a finite closed set of
eigenoperators of the Hamiltonian and exact solution can be
obtained. The purpose of this letter is to define this class, and
to illustrate through some examples the technique of calculations.
Details of calculations and further results will be given
elsewhere.

We consider a system of $q$ species of particles, satisfying Fermi
statistics, subject to finite-range interactions, localized on the
sites of a Bravais lattice. We suppose that the mass of the
particles is very large and/or the interaction is so strong that the
kinetic energy is negligible and the particles are frozen on the
lattice sites. For a two-body interaction the Hamiltonian takes the
form
\begin{equation}
H=\sum_{\mathbf{i}a} V_{a}n_{a}(i) + \frac{1}{2}\sum_{\mathbf{ij}}
\sum_{ab}V_{ab}(\mathbf{i,j})n_{a}(i)n_{b}(j)\label{1}
\end{equation}
where $\mathbf{i}$ stands for the lattice vector $\mathbf{R}_{i}$
and $i=(\mathbf{i},t)$. $c_{a}(i)$ and $c_{a}^{\dag}(i)$ are
annihilation and creation operators of particles of species $a$,
in the Heisenberg picture, satisfying canonical anti-commutation
relations.
\begin{equation}
\begin{array}
[c]{l}
\{c_{a}(\mathbf{i},t),c_{b}^{\dag}(\mathbf{j},t)\}=\delta_{ab}\delta
_{\mathbf{ij}}\\
\{c_{a}(\mathbf{i},t),c_{b}(\mathbf{j},t)\}=\{c_{a}^{\dag}(\mathbf{i}
,t),c_{b}^{\dag}(\mathbf{j},t)\}=0
\end{array}
\label{2}
\end{equation}
$V_{a}$ represents an external field acting on the particles $a$.
$n_{a}(i)=c_{a}^{\dag}(i)c_{a}(i)$ is the charge density operator
of the fields $a$. $V_{ab}(\mathbf{i,j})$ describes the two-body
interactions. It is immediate to see that the $n_{a}(i)$ satisfies
the equation of motion i$\frac{\partial n_{a}(i)}{\partial
t}=[n_{a}(i),H]=0$. Then, in order to use the equations of motion
and Green's function (GF) formalism we must start from the
Heisenberg equation for the $a$ species
\begin{equation}
\mathrm{i}\frac{\partial c_{a}(i)}{\partial t}=V_{a}c_{a}(i)+\sum_
{\mathbf{j}b} V_{ab}(\mathbf{i,j})n_{b}(j)c_{a}(i)\label{3}
\end{equation}
where, without loss of generality, we put $V_{aa}(\mathbf{i,i})=0$.
We see that the dynamics has generated another field operator
$\varphi_{a} (i)=\sum_{\mathbf{j}b}
V_{ab}(\mathbf{i,j})n_{b}(j)c_{a}(i)$ of higher complexity. By
taking time derivatives of increasing order, more and more complex
operators are generated. These operators are composite operators
\cite{Mancini03,Mancini04}, as they are all expressed in terms of
the original fields $c_{a}(i)$ and $c_{a}^{\dag}(i)$. Now, I assert
that: \textbf{because of the algebra satisfied by the charge density
operator }$n_{a}(i)$\textbf{, the hierarchy of composite operators
closes}. The proof of the statement is the following. For the sake
of simplicity, let us take
$V_{ab}(\mathbf{i,j})=2d\delta_{ab}V\alpha_{\mathbf{ij}}$, where $d$
is the dimensionality of the system, $\alpha_{\mathbf{ij}}$ is the
projector on the nearest-neighbor sites and $V$ is the strength of
the interaction. I am considering systems with first-nearest
neighbor interactions, but the following proof can be extended to
systems with second, third, $\cdots$-nearest neighbor interactions.
By increasing the range of interactions the number of eigenoperators
will increase. The time derivative of order $k$ is
\[
\left(  \mathrm{i}\frac{\partial}{\partial t}\right)
^{k}c_{a}(i)=V_{a} ^{k}c_{a}(i)+\sum_{m=1}^{k}mV_{a}^{k-m}\left(
2dV\right) ^{m}\left[ n^{\alpha}(i)\right]  ^{m}c_{a}(i)
\]
where
$n^{\alpha}(i)=\sum_{\mathbf{j}b}\alpha_{\mathbf{ij}}n_{b}(j)$.
Because of (\ref{2}), the operator $\left[  n^{\alpha}(i)\right]
^{m}$ obeys the following recurrence rule

\begin{equation}
\lbrack n^{\alpha}(i)]^{m}=\sum_{p=1}^{2qd}A_{p}
^{(m)}[n^{\alpha}(i)]^{p}\;\;\;\;\;\;\;\sum_{p=1}^{2qd}
A_{p}^{(m)}=1\label{5}
\end{equation}
The coefficients $A_{p}^{(m)}$ are rational numbers that can be
easily determined by the algebra and the structure of the lattice.
Then, for $k>2qd$ no additional composite operators are generated
and the equations of motion close. By putting $n=2qd+1$, we
construct a $n$-multiplet composite operator
\begin{equation}
\psi(i)=\left(
\begin{array}
[c]{c}
\psi_{1}(i)\\
\psi_{2}(i)\\
\vdots\\
\psi_{n}(i)
\end{array}
\right) \label{8}
\end{equation}
which satisfies the Heisenberg equation
\begin{equation}
\mathrm{i}\frac{\partial\psi(i)}{\partial
t}=[\psi(i),H]=\epsilon\psi (i)\label{9}
\end{equation}
where the $n\times n$ matrix $\epsilon$ will be denominated as the
energy matrix. Note that we are using a vectorial notation: the
field $\psi_{m}(i)$ is itself a multiplet of rank $q$. Once the
composite operator $\psi(i)$ and the energy matrix have been
determined, an exact solution of the Hamiltonian can be obtained.
Let us define the retarded Green's function
\begin{equation}
G(i,j)=\theta(t_{i}-t_{j})\left\langle
\{\psi(i),\psi^{\dag}(j)\}\right\rangle \label{10}
\end{equation}
where $\left\langle  \cdots\right\rangle $ denotes the
quantum-statistical average over the grand canonical ensemble. By
means of the Heisenberg equation (\ref{9}) we obtain in momentum
space the equation
\begin{equation}
\lbrack\omega-\epsilon]G(\mathbf{k},\omega)=I(\mathbf{k})\label{12}
\end{equation}
where $I(\mathbf{k})$ is the Fourier transform of the normalization
matrix: $I(\mathbf{i},\mathbf{j})=\left\langle
\{\psi(\mathbf{i},t),\psi^{\dag}(\mathbf{j} ,t)\}\right\rangle $.
The solution of Eq. (\ref{12}) is
\begin{equation}
G(\mathbf{k},\omega)=\sum_{m=1}^n\frac{\sigma
^{(m)}(\mathbf{k})}{\omega-E_{m}+\mathrm{i}\delta}\label{14}
\end{equation}
where $E_{m}$ are the eigenvalues of the energy matrix $\epsilon$.
The spectral density matrices $\sigma^{(m)}(\mathbf{k})$ are
calculated by means of the formula \cite{Mancini03}
\begin{equation}
\sigma_{ab}^{(m)}(\mathbf{k})=\Omega_{am}\sum_{c=1}^{n}\Omega_{am}^{-1}I_{cb}(\mathbf{k})\label{15}
\end{equation}
where $\Omega$ is the $n\times n$ matrix whose columns are the
eigenvectors of the matrix $\epsilon$. The spectral density
matrices $\sigma^{(m)} (\mathbf{k})$ satisfy the sum rule
$\sum_{m=1}^n
E_{m}^{p}\sigma^{(m)}(\mathbf{k})=M^{(p)}(\mathbf{k})$, where
$M^{(p)}(\mathbf{k})$ are the spectral moments defined as
\begin{equation}
M^{(p)}(\mathbf{k})=F.T.\left\langle  \{(\mathrm{i}\partial/\partial
t)^{p} \psi(\mathbf{i},t),\psi^{\dag}(\mathbf{j},t)\}\right\rangle
\label{17}
\end{equation}
$F.T.$ stays for the Fourier transform. The correlation function
(CF) $C(i,j)=\left\langle  \psi(i)\psi^{\dag}(j)\right\rangle $
can be immediately calculated from (\ref{14}) and one obtains
\begin{equation}
C(\mathbf{k},\omega)=\pi \sum_{m=1}^n \delta[\omega
-E_{m}]T_{m}\sigma^{(m)}(\mathbf{k})\label{18}
\end{equation}
with $T_{m}=1+\tanh(\beta E_{m}/2)$ and $\beta=1/k_{B}T$ is the
inverse temperature.

Equations (\ref{14}) and (\ref{18}) are an exact solution of the
model Hamiltonian (\ref{1}). However, the knowledge of the GF is
not fully achieved yet. The algebra of the field $\psi(i)$ is not
canonical: as a consequence, the normalization matrix
$I(\mathbf{k})$ in the equation of motion (\ref{12}) contains some
unknown static correlation functions. These correlators are
expectation values of operators not belonging to the chosen basis
$\psi(i)$, and should be self-consistently calculated.

According to the scheme of calculations proposed by COM
\cite{Mancini03,Mancini03a,Mancini04}, one way of calculating the
unknown correlators is by specifying the representation where the
GF is realized. The procedure is the following. From the algebra
it is possible to derive several relations among the operators. We
will call algebra constraints (AC) all possible relations among
the operators dictated by the algebra. This set of relations valid
at microscopic level must be satisfied also at macroscopic level
(i.e., when expectations values are considered). Use of these
considerations leads to some self-consistent equations which will
be used to fix the unknown correlator appearing in the
normalization matrix. An immediate set of rules is given by the
equation
\begin{equation}
\left\langle  \psi(i)\psi^{\dag}(i)\right\rangle =\frac{1}{N}
\sum_{\mathbf{k}}\frac{\mathrm{1}}{2\pi}\int_{-\infty}^{+\infty}d\omega
C(\mathbf{k},\omega)\label{19}
\end{equation}
where the l.h.s. is fixed by the AC and the boundary conditions
compatible with the phase under investigation, while in the r.h.s.
the correlation function $\left\langle
\psi(i)\psi^{\dag}(i)\right\rangle $ is computed by means of
(\ref{18}).

Another set of AC can be derived by observing that there exist
some operators, O, which project out of the Hamiltonian a reduced
part $OH=OH_{0}$. When $H_{0}$ and $H_{I}=H-H_{0}$ commute, the
quantum statistical averages of O over the complete Hamiltonian
must coincide with the average over the reduced Hamiltonian
$H_{0}$
\begin{equation}
Tr\{Oe^{-\beta H}\}=Tr\{Oe^{-\beta H_{0}}\}\label{21}
\end{equation}

Another relation comes from the requirement of time translational
invariance which leads to the condition that the spectral moments,
defined by Eq. (\ref{17}), must satisfy the following relation
\begin{equation}
M_{nm}^{(p)}(\mathbf{k})=M_{mn}^{(p)}(\mathbf{k})^{*}\label{22}
\end{equation}
It can be shown that if (\ref{22}) is violated, states with
negative norm appear in the Hilbert space. Of course, the above
rules are not meant to be exhaustive and, in principle, more
conditions might be needed.

We now apply the above procedure to two specific models. It goes
without saying that the proposed models belong to the class of the
Potts models \cite{Potts52}. We consider a $d$-dimensional cubic
Bravais lattice.

\textbf{The two-state model}

As first example, we consider only one species of particles. By
taking $V_{a}=-\mu$,
$V_{ab}(\mathbf{i,j})=2dV\delta_{ab}\alpha_{\mathbf{i,j}}$ the
Hamiltonian (\ref{1}) takes the form
\begin{equation}
H=-\mu\sum_{\mathbf{i}} v(i)+Vd \sum_{\mathbf{i}}
v(i)v^{\alpha}(i)\label{25}
\end{equation}
where hereafter for a generic operator $\Phi(i)$ we use the
notation
$\Phi^{\alpha}(\mathbf{i},t)=\sum_{\mathbf{j}}\alpha_{\mathbf{ij}
}\Phi(\mathbf{j},t)$. $\mu$ is the chemical potential and we put
$v(i)=c^{\dag}(i)c(i)$; we are ignoring the subindex $a$. By means
of the transformation $v(i)=\frac{1}{2}[1+S(i)]$ the Hamiltonian
(\ref{25}) can be cast in the form
\begin{equation}
H=E_{0}-h\sum_{\mathbf{i}} S(i)-dJ\sum_{\mathbf{i}}
S(i)S^{\alpha}(i)\label{28}
\end{equation}
where $E_{0}=(-\frac{1}{2}\mu+\frac{1}{4}dV)N$,
$h=\frac{1}{2}(\mu-dV)$, $J=-\frac{1}{4}dV$. Hamiltonian
(\ref{28}) is just the $d$-dimensional spin-1/2 Ising model
\cite{Ising25} with nearest neighbor interactions in presence of
an uniform external magnetic field.

To solve the Hamiltonian (\ref{25}) let us consider the composite
operator\vspace{-0.1cm}

\begin{equation}
\psi(i)=\left(
\begin{array}
[c]{c}
\psi_{1}(i)\\
\psi_{2}(i)\\
\vdots\\
\psi_{2d+1}(i)
\end{array}
\right)  =\left(
\begin{array}
[c]{c}
c(i)\\
v^{\alpha}(i)c(i)\\
\vdots\\
\lbrack v^{\alpha}(i)]^{2d}c(i)
\end{array}
\right) \label{30}
\end{equation}
This field is an eigenoperator of the Hamiltonian (\ref{25})
\begin{equation}
\mathrm{i}\frac{\partial}{\partial
t}\psi(i)=[\psi(i),H]=\epsilon\psi (i)\label{36}
\end{equation}
The energy matrix $\epsilon$ has the rank $(2d+1)\times(2d+1)$ and
can be calculated by means of (\ref{3}) and the recurrence rule
(\ref{5}). Proof of the relation (\ref{5}) and the explicit
expressions of $A_{m}^{(p)}$ for this model will be given
elsewhere. The eigenvalues of the energy matrix are given by
\begin{equation}
E_{m}=-\mu+(m-1)V\;\;\;\;\;\;\{m=1,2,\cdots\cdots(2d+1)\}\label{37}
\end{equation}

The retarded GF $G(i,j)$ and the CF $C(i,j)$ can be exactly
calculated by applying the scheme of calculations illustrated
above. By using the anti-commutation rules (\ref{2}) and the
symmetry relations (\ref{22}), we obtain
$I(\mathbf{i,j})=\delta_{\mathbf{i,j}}I_{0}$, where the matrix
$I_{0}$ depends on the parameters $\kappa^{(p)}=\langle
[v^{\alpha}(i)]^{p}\rangle $ $(p=1,\cdots 2d)$. Then, the matrices
$\sigma^{(m)}$ are calculated by means of (\ref{15}). We have an
exact solution of the Ising model for dimensions $d=1,2,3$, where
all the properties of the model are expressed in terms of the
self-consistent parameters $\kappa^{(p)}$.\ To have quantitative
results we must calculate $\kappa^{(p)}$ in terms of the external
parameters $\mu,T$ and $V$. We now recall the AC (\ref{19}).
Because of the relation (\ref{5}) there are only $2d+1$
independent elements of the matrix $C=\langle
\psi(i)\psi^{\dag}(i)\rangle $ and we obtain the self-consistent
equations
\begin{equation}
\kappa^{(p)}-\lambda^{(p)}=\frac{1}{2} \sum_{m=1}^{2d+1}
T_{m}\sigma_{1,p}^{(n)}\;\;\;\;\;\;(p=1,\cdots2d+1)\label{38}
\end{equation}
where $T_{m}=1+\tanh(\beta E_{m}/2)$ and $\sigma^{(n)}$ are the
spectral density matrices. New correlation functions
$\lambda^{(p)}=\langle v(i)[v^{\alpha }(i)]^{p}\rangle $ appear
and the set of self-consistent equations (\ref{38}) is not
sufficient to determine all unknown parameters. One needs more
conditions. In the case $d=1$, these extra conditions can be
obtained by using property (\ref{21}). By means of the algebraic
relation $c^{\dag}(i)v(i)=0$, we have $c^{\dag}(i)e^{-\beta
H}=c^{\dag}(i)e^{-\beta H_{0}}$, where $H_{0}
=H-2Vv(i)v^{\alpha}(i)$. By requiring that (\ref{21}) be
satisfied, we can derive the relation
$\frac{C_{1k}}{C_{11}}=\frac{C_{1k}^{(0)}}{C_{11}^{(0)}}$ where
$C_{1k}=\left\langle
c(i)c^{\dag}(i)[v^{\alpha}(i)]^{k-1}\right\rangle $ and
$C_{1k}^{(0)}=\left\langle  c(i)c^{\dag}(i)[v^{\alpha}(i)]^{k-1}
\right\rangle _{0}$. Here $\left\langle  \cdots\right\rangle _{0}$
denotes the thermal average with respect to $H_{0}$. By exploiting
these relations we can arrive to the following equation
\begin{equation}
C_{13}=\frac{1}{2}C_{12}\left(  1+\frac{C_{12}}{C_{11}}\right)
\label{45}
\end{equation}
This equation closes the system of self-consistent equations and
we have a complete solution of the model. Higher-order correlation
functions can be studied by similar techniques. All details of the
calculations will be given in a separate work, where we will show
that all the well-known results
\cite{Baxter82,Goldenfeld92,Lavis99} of the one-dimensional Ising
model are recovered.

\textbf{Three-state model}

We consider now a second example. Let us consider two species of
particles, say $a$ and $b$, and take $V_{a}=V_{b}=-\mu$,
$V_{ab}(\mathbf{i,j}
)=U\delta_{\mathbf{ij}}\delta_{ab}+2dV\alpha_{\mathbf{ij}}$. Then,
the Hamiltonian (\ref{1}) becomes
\begin{equation}
H=\sum_{\mathbf{i}} [-\mu n(i)+UD(i)+Vn(i)n^{\alpha}(i)]\label{52}
\end{equation}
where $n(i)=n_{a}(i)+n_{b}(i)$ and $D(i)=n_{a}(i)n_{b}(i)=\frac{1}
{2}n(i)[n(i)-1]$ are the total particle density and double occupancy
operators, respectively. This Hamiltonian is just the extended
Hubbard model in the ionic limit, where $U$ and $V$ are the on-site
and inter-site Coulomb interaction, respectively. The two species of
particles, $a$ and $b$, are in this case electrons with spin up and
down, respectively. By means of the transformation $n(i)=[1+S(i)]$,
(\ref{52}) can be cast in the form
\begin{equation}
H=-dJ\sum_{\mathbf{i}} S(i)S^{\alpha}(i)+\Delta \sum_{\mathbf{i}}
S^{2}(i)-h \sum_{\mathbf{i}} S(i)+E_{0} \label{54b}
\end{equation}
where $E_{0}=(-\mu+dV)N$, $h=\mu-2dV-\frac{1}{2}U$, $J=-dV$,
$\Delta=\frac {1}{2}U$. Hamiltonian (\ref{54b}) is just the Ising
spin-1 model \cite{Blume71} with nearest-neighbor interactions in
presence of a crystal field $\Delta$ and an external magnetic
field $h$.

To solve the Hamiltonian (\ref{52}) let us consider the composite
operators
\begin{equation}
\psi^{(x)}(i)=\left(
\begin{array}
[c]{c}
\psi_{1}^{(x)}(i)\\
\psi_{2}^{(x)}(i)\\
\vdots\\
\psi_{4d+1}^{(x)}(i)
\end{array}
\right)  =\left(
\begin{array}
[c]{c}
x(i)\\
x(i)[n^{\alpha}(i)]\\
\vdots\\
x(i)[n^{\alpha}(i)]^{4d}
\end{array}
\right) \label{62}
\end{equation}
where $x=\xi,\eta,$ being $\xi(i)=[1-n(i)]c(i)$ and
$\eta(i)=n(i)c(i)$ the Hubbard operators.\ Note that $c(i),$
$\xi(i),\eta(i)$ are doublet operators with components $a$ and
$b$. By means of (\ref{3}) and (\ref{5}), these fields are
eigenoperators of the Hamiltonian (\ref{52})
\begin{equation}
\begin{array}
[c]{l} \mathrm{i}\frac{\partial}{\partial
t}\psi^{(\xi)}(i)=[\psi^{(\xi)}
(i),H]=\epsilon^{(\xi)}\psi^{(\xi)}(i)\\
\mathrm{i}\frac{\partial}{\partial t}\psi^{(\eta)}(i)=[\psi^{(\eta
)}(i),H]=\epsilon^{(\eta)}\psi^{(\eta)}(i)
\end{array}
\label{63}
\end{equation}
where $\epsilon^{(\xi)}$ and $\epsilon^{(\eta)}$ are the energy
matrices, of rank $(4d+1)\times(4d+1)$, which can be calculated by
means of the equations of motion (\ref{63}) and the recurrence
relation (\ref{5}). The eigenvalues $E_{m}^{(\xi)}$ and
$E_{m}^{(\eta)}$ of the energy matrices are
\[
\begin{array}
[c]{l}
E_{m}^{(\xi)}=-\mu+(m-1)V\\
E_{m}^{(\eta)}=-\mu+U+(m-1)V
\end{array}
\;\;\;\;\;\;\{m=1,2,\cdots(4d+1)\}
\]

The retarded GF $G^{(xy)}(i,j)=\left\langle
R[\psi^{(x)}(i)\psi^{(y)\dag }(j)]\right\rangle $ and the CF
$C^{(xy)}(i,j)=\left\langle  \psi^{(x)}
(i)\psi^{(y)\dag}(j)\right\rangle $, with $x,y=\xi,\eta$, can be
exactly calculated by applying the scheme of calculations
illustrated above. By using the anti-commutation relations
(\ref{3}) and the symmetry relations (\ref{22}), straightforward
calculations show that $I^{(\xi\eta)}
(\mathbf{i,j})=I^{(\eta\xi)}(\mathbf{i,j})=0$ , while
\[
\begin{array}
[c]{l}
I_{1,k}^{(\xi\xi)}(\mathbf{i,j})=\delta_{\mathbf{ij}}[\kappa^{(k-1)}
-\lambda^{(k-1)}]\\
I_{1,k}^{(\eta\eta)}(\mathbf{i,j})=\delta_{\mathbf{ij}}\lambda^{(k-1)}
\end{array}
\;\;\;\;(k=1,\cdots,4d+1)
\]
where $\kappa^{(p)}=\langle [n^{\alpha}(i)]^{p}\rangle $,
$\lambda^{(p)}=\frac{1} {2}\langle n(i)[n^{\alpha}(i)]^{p}\rangle
$.\ The CF $C^{(xx)}(i,j)$, $(x=\xi,\eta)$, have the expressions

\begin{equation}
C^{(xx)}(i,j)=\delta_{\mathbf{ij}}\frac{1}{2}\sum_{m=1}^{4d+1}
T_{m}^{(x)}\sigma^{(xm)}e^{-iE_{m}^{(x)}(t_{i}-t_{j})}\label{68}
\end{equation}
where $T_{m}^{(x)}=1+\tanh(\beta E_{m}^{(x)}/2)$. The spectral
density matrices $\sigma^{(xm)}$ are expressed in terms of the
elements $I_{1k} ^{(xx)}$. We have an exact solution of the model
for dimensions $d=1,2,3$, where all the properties are expressed
in terms of the self-consistent parameters $\kappa^{(p)}$and
$\lambda^{(p)}$.\ In order to determine the correlators
$\kappa^{(p)}$ and $\lambda^{(p)}$, we use the AC constraints
(\ref{19}). By means of the algebraic relations $\xi_{c}(i)\xi
_{c}^{\dag}(i)+\eta_{c}(i)\eta_{c}^{\dag}(i)=1-n_{c}(i)$, with
$c=a,b$, we obtain for the paramagnetic phase the following
self-consistent equations
\begin{equation}
\kappa^{(k-1)}-\lambda^{(k-1)}=C_{1k}^{(\xi\xi)}+C_{1k}^{(\eta\eta)}\quad
\quad(k=1,\cdots4d+1)\label{72}
\end{equation}
The number of equations is not sufficient to fix all the
parameters and more equations are needed. In the case of $d=1$
these extra conditions can be obtained by using the property
(\ref{21}). By means of the algebraic relations
\begin{equation}
\begin{array}
[c]{l}
{{\xi^{\dag}(i)n(i)=0}}\\
{{\xi^{\dag}(i)D(i)=0}}
\end{array}
\qquad
\begin{array}
[c]{l}
{D}^{p}{(i)=D(i)}\\
{D(i)n}^{p}(i){=2}^{p}D(i)
\end{array}
\label{73}
\end{equation}
we have
\begin{equation}
\begin{array}
[c]{l}
{{\xi^{\dag}(i)e}}^{{-\beta H}}{{=\xi^{\dag}(i)e}}^{{-\beta H}_{0}}\\
D(i)e^{-\beta
H}=D(i)\sum\limits_{p=0}^{4}{}h_{p}[n^{\alpha}(i)]^{p}e^{-\beta
H_{0}}
\end{array}
\label{74}
\end{equation}
where $H_{0}=H-2Vn(i)n^{\alpha}(i)$, and the $h_{p}^{\prime}s$ are
known functions of the potential $V$. By requiring that (\ref{21})
be satisfied we can derive a set of three self-consistent
equations, which added to (\ref{72}) allow us to determine all 8
parameters $\kappa^{(p)}$and $\lambda^{(p)}$ $(p=1,\cdots4)$.
Calculations and results will be presented elsewhere.

Summarizing, we have shown that a system of localized particles,
satisfying Fermi statistics, subject to finite-range interactions,
can be described in terms of a closed set of eigenoperators. For the
case of nearest neighbor interactions, the number of these composite
operators is equal to $(2qd+1)$, where $q$ is the number of species
of particles and $d=1$, $2$, $3$ is the dimensionality of the
system. The Green's functions and the correlation functions can be
exactly calculated and are expressed in terms of a set of
self-consistent parameters. For example, we have considered two
models: the two-state model (equivalent to the spin-$\frac{1}{2}$
Ising model) and the three-state model (equivalent to the extended
ionic Hubbard model and to the spin-$1$ Ising model). For these
models the parameters are calculated in the case of $d=1$. For
higher dimensions more self-consistent equations are needed. This
problem is now under investigation.

\acknowledgments The author is grateful to Dr.~A.~Avella for
stimulating discussions, a very friendly collaboration and his
careful reading of the manuscript.


\newcommand{\noopsort}[1]{} \newcommand{\printfirst}[2]{#1}
  \newcommand{\singleletter}[1]{#1} \newcommand{\switchargs}[2]{#2#1}

\end{document}